\documentclass[twocolumn,amsmath,superscriptaddress]{revtex4-1}

\usepackage{graphics}
\usepackage{amssymb,amsfonts,amsmath}
\usepackage{mleftright}
\usepackage{xparse}
\usepackage{amsfonts}
\usepackage{mathtools}
\usepackage{physics}
\usepackage[svgnames]{xcolor}

\usepackage{hyperref}

    \usepackage{graphicx} 
\usepackage[figurename=Figure]{caption}
\usepackage{amsmath}  
\usepackage{amsthm}  
\usepackage{amssymb}  

\def \>{\rangle}
\def \<{\langle}

\def\be{\begin{equation}}
\def\ee{\end{equation}}
\def\longrightharpoonup{\relbar\joinrel\rightharpoonup}
\def\longleftharpoondown{\leftharpoondown\joinrel\relbar}

\def\longrightleftharpoons{
  \mathop{
    \vcenter{
      \hbox{
      \ooalign{
        \raise1pt\hbox{$\longrightharpoonup\joinrel$}\crcr
	  \lower1pt\hbox{$\longleftharpoondown\joinrel$}
	  }
      }
    }
  }
}

\newcommand \bea {\begin{eqnarray}}
\newcommand \eea {\end{eqnarray}}

\NewDocumentCommand{\evalat}{sO{\big}mm}{%
  \IfBooleanTF{#1}
   {\mleft. #3 \mright|_{#4}}
   {#3#2|_{#4}}%
}

\theoremstyle{definition}

\newcommand{\R}{\mathbb R}

\begin{document}
\title{ Geometry of ecological coexistence and niche differentiation}

\author{Emmy Blumenthal}
\affiliation{Department of Physics, Boston University, Boston, Massachusetts 02215, USA}
\author{Pankaj Mehta}
\affiliation{Department of Physics, Boston University, Boston, Massachusetts 02215, USA}
\affiliation{Biological Design Center, Boston University, Boston, Massachusetts 02215, USA}
\affiliation{Faculty of Computing and Data Sciences, Boston University, Boston, Massachusetts 02215, USA}
 \begin{abstract}
        A fundamental problem in ecology is to understand how competition shapes biodiversity and species coexistence.  Historically, one important approach for addressing this question has been to analyze consumer resource models using geometric arguments. This has led to broadly applicable principles such as Tilman's $R^*$ and species coexistence cones. Here, we extend these arguments by constructing a novel geometric framework for understanding species coexistence based on convex polytopes in the space of consumer preferences. We show how the geometry of consumer preferences can be used to predict species which may coexist and enumerate ecologically-stable steady states and transitions between them.
        Collectively, these results provide a framework for understanding the role of species traits within niche theory.
     \end{abstract}
     
\maketitle

\section{Introduction}

One of the most striking features of natural ecosystems is the immense diversity of flora and fauna they support. Understanding the origin of this diversity is a fundamental question in ecology. A single ecosystem can contain  thousands of species, all living in close proximity and interacting with each other and the abiotic environment. For this reason, theoretical models have played a major role in guiding empirical research, helping to interpret experiments, and shaping ecological intuitions \cite{chase2009ecological, tilman1982resource, levins2020evolution}.

One major theoretical framework for understanding biodiversity is niche theory \cite{chase2009ecological}. Niche theory emphasizes the central role played by competition in shaping ecological properties. Within the niche paradigm,  species must occupy distinct niches to coexist in an ecosystem \cite{macarthur1967limiting, armstrong1980competitive}. This idea is most succinctly summarized in Tilman's R* principle which states that every surviving species must be best at utilizing a different resource \cite{tilman1982resource}. Within the niche theory, the biodiversity of an ecosystem is determined by  the number of distinct niches that species can occupy \cite{macarthur1969species, levin1970community}.

\begin{figure}[t]
    \includegraphics[width=\linewidth]{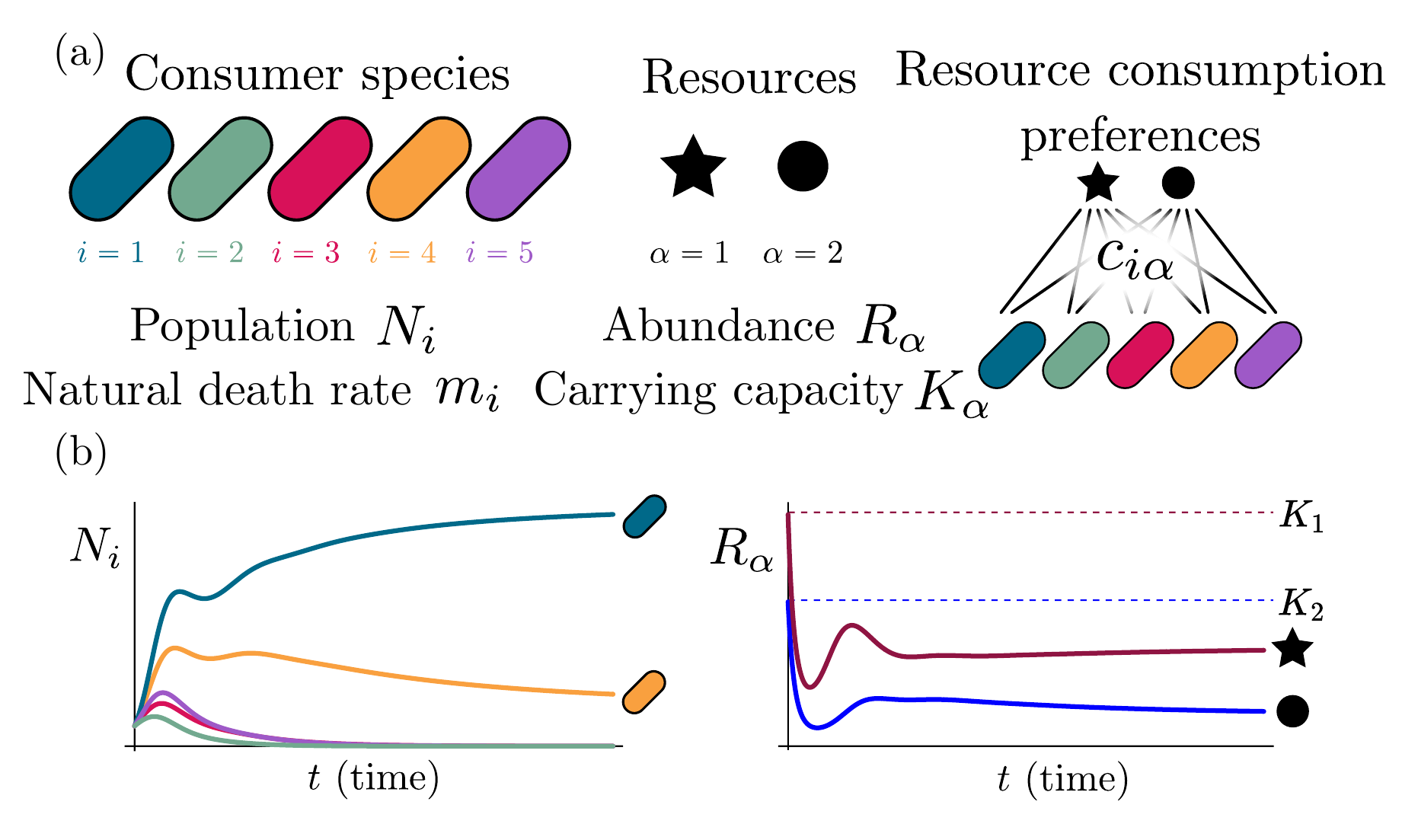}
    \caption{MacArthur consumer resource model (a) A species $i$ consumes resource $\alpha$ with preference set by consumer preferences $c_{i \alpha}$. (b) Population dynamics for a system with two resources and four species described by Eq. \ref{Eq:CRM}. (see the Appendix~\ref{appendix:TwoResourceModel} for parameters)}
  \label{fig:CRM}
\end{figure}

Consumer resource models (CRMs) have played a central role in the development of niche theory (Fig.~\ref{fig:CRM}) \cite{macarthur1970species, macarthur1969species, chesson1990macarthur}. CRMs consists of two kinds of variables: resources and consumers. Each consumer is defined by a set of ``consumer preferences'' indicating which resources it can utilize.
Species-species interactions arise through competition for the common pool of resources. An appealing feature of CRMs is that they explicitly model both species and resources, allowing us to understand environmental conditions in which species can coexist and when they competitively exclude each other.

Recently, there has been a renewed interest in consumer resource models from both the ecology and statistical physics communities. A number of works have analyzed these models using methods from statistical physics to understand the behavior of complex ecosystems with many species and resources \cite{tikhonov2017collective, cui2020effect, cui2021diverse}. Other works have generalized these models to study microbial ecology by  incorporating metabolic considerations \cite{goldford2018emergent, marsland2018available, niehaus2019microbial, muscarella2020species, bajic2020ecology, gowda2022genomic}, including the role of metabolic tradeoffs \cite{d2020emergent, posfai2017metabolic, li2020modeling}. CRMs are also being increasingly used as a test-ground for furthering our understanding of eco-evolutionary dynamics \cite{good2018adaptation, caetano2021evolution, tikhonov2020model, goyal2022interactions, chang2021engineering} and community selection \cite{xie2019simulations, chang2021engineering, sanchez2022community}.
 
Historically, many of the central intuitions of niche theory have been developed using geometric arguments for analyzing CRMs \cite{leibold1996graphical,levins1966strategy}. For example, Tilman's R* principle encodes the conditions for species coexistence as geometric intuitions about when zero net-growth isoclines (ZNGIs) of different species  -- defined as the set of resource abundances for which a species has zero growth rate -- intersect in resource abundance space (i.e., the vector space of possible resource abundances) \cite{tilman1982resource}. Geometric reasoning is also fundamental to the development of contemporary niche theory \cite{chase2009ecological}. Within contemporary niche theory,  whether two species can coexist is determined by the geometry of  ``coexistence cones''  in resource abundance space (Fig.~\ref{fig:oldPicture}).

Traditional geometric arguments for determining species coexistence primarily work in resource abundance space. In this work, we derive an alternative geometric representation of species coexistence and niche differentiation in CRMs using a convex polytope in the space of ``species consumption preferences.''
This geometric analysis extends previous work by foregrounding the central role played by niche differentiation in coexistence. It also allows us to enumerate all possible ecologically stable steady states, as well as all transitions between steady states due to changes in resource supply. 
For this reason, it represents an interesting new way of understanding the origin of biodiversity within the context of niche theory.

\section{ Methods and Results}

\subsection{Model}
To illustrate our ideas, we initially focus on MacArthur's consumer resource model (MCRM) \cite{macarthur1970species, macarthur1969species, chesson1990macarthur}.
In Sec.~\ref{sec:other-models}, we show how this analysis can be extended to variants of the consumer resource model (CRMs) including those analyzed by Tilman \cite{tilman2020resource}, CRMs with alternative resource dynamics \cite{cui2019effect}, and CRMs with nonlinear growth rates.

The MCRM consists of $S$ species (also called  consumers) with abundances $N_i$ ($i=1,\ldots, S$) that can consume $M$ resources with abundances $R_\alpha$ ($\alpha=1,\ldots, M$).
Species grow by consuming resources and die at a per-capita rate $m_i$.
Alternatively, $m_i$ can be interpreted as the minimum amount of energy that must be extracted from resources in order for species $i$ to survive. Resources also have different ``qualities''  $w_\alpha$ reflecting how much they contribute to growth. The preference of species $i$ for a resource $\alpha$ is encoded in the matrix of consumption preferences $c_{i \alpha}$. In the absence of consumers, each resource is described by logistic growth with carrying capacity $K_\alpha$. In the presence of consumers, resources are depleted at a rate proportional to their consumption. 
 
The dynamics of the MCRM are described by the following coupled ordinary differential equations:
\begin{align}
    \frac{\mathrm dR_\alpha}{\mathrm dt}&=R_\alpha \left(K_\alpha - R_\alpha
    \right)-\sum_{i=1}^S c_{i\alpha} N_i R_\alpha \nonumber \\
   \frac{\mathrm dN_i}{\mathrm dt} &=N_i \left(\sum_{\alpha = 1}^M w_\alpha c_{i\alpha} R_\alpha-m_i \right).
     \label{Eq:CRM}
\end{align}
In general, at steady state, some resources and species will go extinct. Denote the number of species and resources that survive at steady state by $S^\star$ and $M^\star$, respectively. 

Generically, the principle of competitive exclusion implies that at most $S^\star\leq M^\star$ species can survive in an ecosystem at steady state \cite{armstrong1980competitive} (but also see Refs.~\cite{posfai2017metabolic, cui2019effect, pacciani2020dynamic} for discussion of exceptions and how they might arise).
An example of the resulting dynamics is shown in Fig.~\ref{fig:CRM} for a system with $S=5$ and $M=2$. Notice that the steady-state abundances of the two resources is below their carrying capacities due to depletion, and the number of surviving species $S^\star=2$ is bounded by the number of surviving resources, in this case $M^\star=2$.

In general, which species survive depends on the resource supply point encoded in the ``resource supply vector,'' $\mathbf{K}$. As shown in Fig.~\ref{fig:oldPicture}, by varying the resource supply point, the steady-state of the MCRM has qualitatively different behaviors:  a single species $i=5$ competitively excludes all species, two species $i=1,5$  coexist, species $i=1$ competitively exclude all species, the two species $i=1,4$ coexist, and finally species $i=4$ competitively excludes all other species.
Notice that species $2$ and $3$ always go extinct and species $4$ and $5$ can never coexist for any choice of the carrying capacities $\mathbf{K}$.
Thus, even a simple model with two resource and five species exhibits an extremely rich set of possible steady-state behaviors.

\subsection{Geometry in resource abundance space\label{sec:resource-space-geom}}


\begin{figure*}[t!]
    \includegraphics[width=0.9\linewidth]{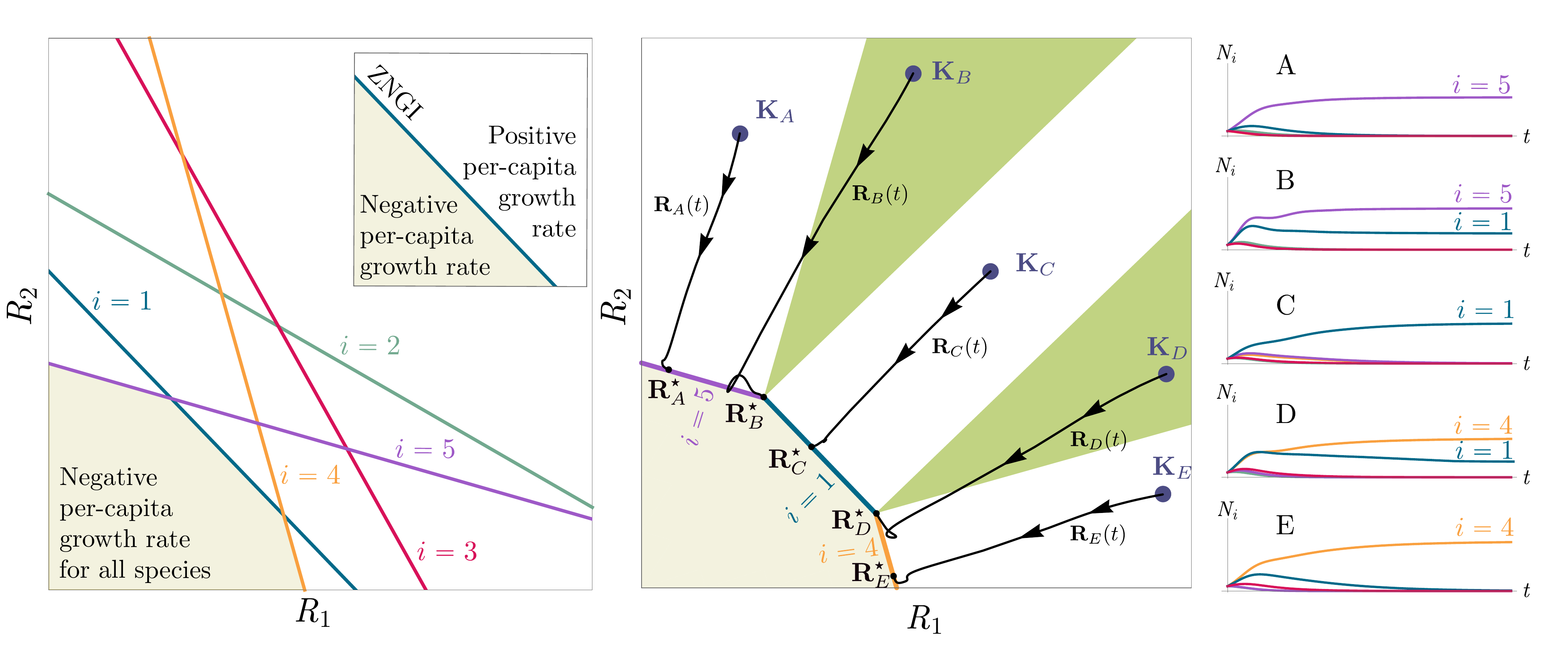}
    \caption{Zero net growth isoclines (ZNGIs) and coexistence cones for ecosystems with $S=5$ species and $M=2$ resources.
    (a) ZNGIs for each species are plotted in the space of resource abundances. The highlighted region corresponds to the polytope defining the infeasible region where the growth rate is negative for all species. (Inset) ZNGI for one species where different growth-rate regions are labeled. (b) Trajectories of solutions in resource phase space (black lines), coexistence cones (green cones), and species abundances $N_i$ as a function of time for five different choices of carrying capacity indicated by $\mathbf{K}_A, \dots, \mathbf{K}_E$. Dynamics in Fig.~\ref{fig:CRM} correspond to trajectory with carrying capacity $\mathbf K_D$. See interactive demonstrations in Appendix~\ref{MMA-notebook-links}.}
     \label{fig:oldPicture}
   \end{figure*}
  
 Contemporary niche theory provides a simple geometric picture for understanding possible steady-state behaviors in CRMs \cite{chase2009ecological}. Consider, once again the simple ecosystem shown in Fig.~\ref{fig:CRM}. A key quantity of interest are the Zero net growth isoclines (ZNGIs). A ZNGI for a given species corresponds the set of resource abundances for which a species has exactly zero growth rate. Setting $\frac{dN_i}{dt} = 0$ in Eq. \ref{Eq:CRM}, we see that the ZNGI for a species $i$ is defined by the equation, 
\be
g_i(\mathbf{R}^\star)=\sum_{\alpha = 1}^M w_\alpha c_{i\alpha} R_\alpha^\star - m_i = 0,
\label{Eq:ZNGI}
\ee
where we have defined the per-capita growth rate $g_i(\mathbf{R}^\star)$ of species $i$ and introduced the notation where $X^\star$ denotes the steady-state value of a quantity $X$. For an ecosystem with $S$ species, we have $S$ different ZNGIs corresponding to the indices $i=1, \ldots, S$.
For each ZNGI, $g_i(\mathbf{R}^\star)=0$ defines a codimension one hyperplane in the space of the resource abundances [Fig.~\ref{fig:oldPicture}(a)].
As shown in the inset, this hyperplane divides the resource space into two regions: a region containing the origin where species $i$ has negative per-capita growth rate and a region where species $i$ has positive per-capita growth rate. 

Equation~\eqref{Eq:CRM} tells us that at steady state, the per-capita growth rate for a species $i$ must either be zero if species $i$ survives (i.e. $g_i(\mathbf{R}^\star)=0$) or negative if species $i$ goes extinct (i.e. $g_i(\mathbf{R}^\star)<0$).
For this reason and because the CRM in Eq. \eqref{Eq:CRM} is guaranteed to reach a steady state for all choices of the carrying capacities $\mathbf{K}$  \cite{mehta2019constrained, marsland2019minimum}, the steady-state resource abundances $\mathbf R^\star$ must lie on the boundary of the convex polytope formed by the intersection of the ZNGIs associated with the surviving species, restricted to the positive quadrant [Fig.~\ref{fig:oldPicture}(a)].
The position of $\mathbf R^\star$ on this boundary depends on the supplied resource vector $\mathbf{K}$.
We emphasize that different choices of $\mathbf{K}$ correspond to \emph{different} ecosystems with distinct distributions of supplied resources (i.e., resource carrying capacities in the absence of consumers).

If we further require that the steady state be \emph{ecologically uninvadable} (i.e., cannot be invaded by any species), then $\mathbf R^\star$ must lie inside the boundary of the polytope formed by the ZNGIs; we refer to this region as the infeasible region.
In Fig.~\ref{fig:oldPicture}(a), this corresponds to the shaded region.
The number and identity of species that coexist is directly related to the number of ZNGIs that intersect at $\mathbf R^\star$ since we must have $g_i(\mathbf{R}^\star)=0$ for all species $i$ that survive at steady state.
For the example in Fig. \ref{fig:oldPicture}, the vertices on the boundary of the shaded region correspond to values of  $\mathbf R^\star$ where two species coexist at steady state, and edges correspond to values of $\mathbf R^\star$ where one species competitively excludes all others.
This basic argument also explains why, generically, the number of surviving species must be less than the number of surviving resources, $S^\star\le M^\star$, because at most $M^\star$ planes can intersect in $M^\star$ dimensions without fine-tuning.

The arrangements of the intersections between the ZNGIs which fall in the intersection of all the closed half-spaces enumerate the possible coexisting species in {\em various} ecosystems, and the dimensions of the intersections correspond dually to the number of coexisting species.
The species that coexist in a {\em particular} ecosystem depend on the resource supply vector $\mathbf K$. At steady state, for a nondepleted resource $\alpha$ Eq.~\eqref{Eq:CRM} states that
\be
 {K}_\alpha- {R}^\star_\alpha=\sum_{i, N_i^\star>0} c_{i \alpha} N_i^\star.
 \label{Eq:cone-eq}
\ee
Geometrically, the left-hand side corresponds to the vector from the steady-state resource values to the resource carrying capacities.
By definition, $N_i^\star >0$, so the right-hand side defines a ``coexistence cone'' with basis given by the consumer resource preferences (Fig.~\ref{fig:oldPicture}(b) and Ref.~\cite{chase2009ecological} for extended discussion).  As shown in the figure,  this equation also implies that $\mathbf{K}$ must lie within the coexistence cones. 

The coexistence cones tile the resource-abundance phase space and classify each $\mathbf K$ by the species whose coexistence it supports.
The number of species that coexist is the rank of the coexistence cone, and the apex of the coexistence cone is the intersection of all ZNGIs corresponding to the surviving species.
In general, if $S^\star$ species survive and coexist, then the coexistence cone has rank $S^\star$, and the intersection of the surviving species' ZNGIs has dimension $M-S^\star$. An interactive {\em Mathematica} notebook illustrating this basic picture can be found on the corresponding GitHub repository (see Appendix~\ref{MMA-notebook-links}).

\subsection{ Geometry in space of consumer preferences}

We now provide an alternative geometric picture of ecosystems that works directly in the space of ``consumer preferences.''
The virtue of working in the space of consumer preferences is that it allows one to directly link species coexistence to species traits, which are often easier to observe and characterize from data.
On a technical level, the geometry in the consumer preference space is mathematically dual to the geometry in resource space and hence contains the same information (see below) \cite{blekherman2012semidefinite}.  However, the qualitative ecological intuitions it provides are quite distinct and meaningful from those discussed in the last section.

To formulate this new geometric picture we introduce ``scaled consumption vectors''  $\mathbf C_i$ for each species:
\begin{align}
  \mathbf C_i
  &= \left(
  \begin{array}{ccccc}
    w_1 c_{i1}/m_i & \dots &
    w_\alpha c_{i\alpha}/m_i & \dots &
    w_M c_{iM}/m_i
  \end{array}
  \right)
  ^\mathrm{T}.
\end{align}
Notice the elements of $\mathbf{C}_i$ are the original consumer preferences for surviving species divided by $m_i$ where resources are weighted by energetic content $w_\alpha$.
In terms of these scaled consumer resources, the steady-state condition $g_i(\mathbf R^\star) \leq 0$ from Eq.~\eqref{Eq:ZNGI} can be rewritten as,
\be
\mathbf C_i \cdot \mathbf R^\star - 1 \leq 0,  
\label{Eq:rescaledZNGI}
\ee
with strict equality for the indices corresponding to the $S^\star$ surviving species. Geometrically, this equation states that the ZNGI for each a species $i$ is
perpendicular to $\mathbf C_i$ and passes through the point $\mathbf C_i/\|\mathbf C_i\|^2$.
This can be easily verified by observing that $\mathbf{R}^\star= \mathbf C_i/\|\mathbf C_i\|^2$ is a solution to Eq.~\eqref{Eq:rescaledZNGI}.

This inversive relationship between the ZNGIs and the consumption vectors allows us to analyze many coexistence properties \emph{using the rescaled consumption vectors alone} without explicit reference to steady-state resource abundances.  In the main text, we limit ourselves to discussing the results and interpretations that follow from this observation and relegate technical details to Appendix \ref{appendix:PCHappendix}.

\begin{figure}[t!]
\includegraphics[width=\linewidth]{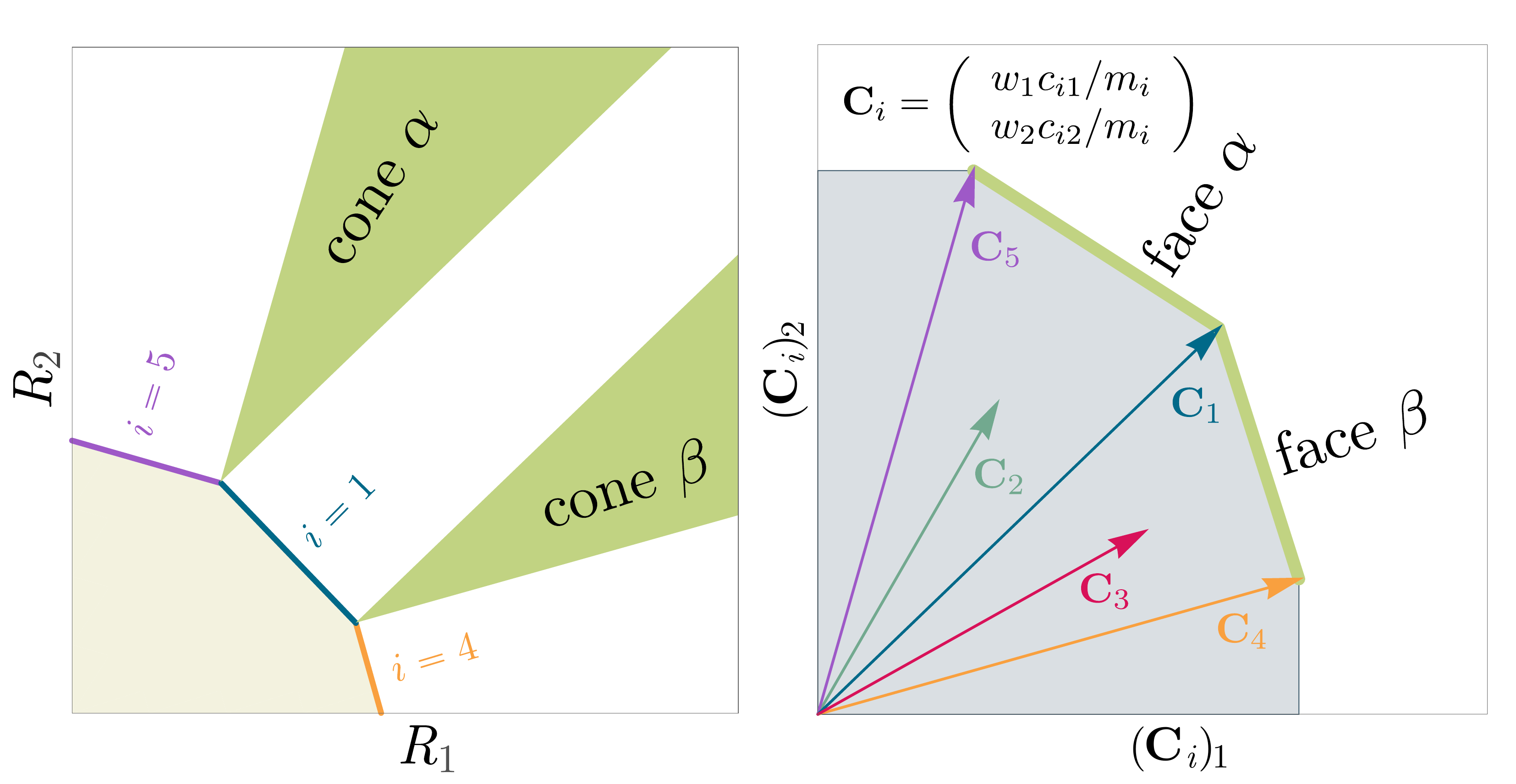}
        \caption[Coexistence cones]{
          Relationship between coexistence cones and the positive convex hull (PCH) formed by rescaled consumption vectors.
            Left: Coexistence cones are highlighted in relation to the ZNGIs for competitive species.
            If the vector of carrying capacities, $\mathbf K$, belongs to cone $\alpha$, the species $i=1, i=5$ can coexist; if $\mathbf K$ belongs to cone $\beta$, then species $i=1,i=4$ can coexist; otherwise, only $i=5,i=1,i=4$ can survive separately.
            Right: 
            The positive convex hull (PCH) of rescaled consumption vectors $\mathbf C_i$ has faces that are dual to the arrangement of the coexistence cones.
            Each possible ecologically stable combination of surviving species is represented by a face of the PCH.
}
\label{fig:PCH}
\end{figure}

The central geometric object in our picture is a convex polytope formed by the scaled consumption preferences $\mathbf{C}_i$ which we call the positive convex hull (PCH) of consumption vectors. 
The PCH is constructed by forming the convex hull of all scaled consumption vectors, $\mathbf C_i$, and these consumption vectors projected onto all combinations of coordinate axes.
It is necessary to include these projections to account for the possibility that resources may be depleted; one may think of these projected vectors as consumption vectors for species that are forbidden from consuming some potentially depleted resources.
Figure \ref{fig:PCH} depicts the PCH corresponding to the ecosystem analyzed in Fig.~\ref{fig:oldPicture}.
The figure also provides an illustration of the geometric duality between the PCH and the convex polytope formed by the ZNGIs in  resource space corresponding to the infeasible region (i.e., the region where growth rates for all species are negative). This duality follows directly from Eq. \ref{Eq:rescaledZNGI} and has a number of powerful implications (see Appendix \ref{appendix:PCHappendix}). Chief among these is that a face of dimension $M-d$ of the infeasible region polytope formed by the ZNGIs is in direct correspondence to a face of dimension $d$ of the PCH.
We use the usual definition that a face is an intersection of a boundary of the polytope with a hyperplane; a face is not necessarily maximal-rank and can be a single vertex.

In Appendix~\ref{justification}, under the same very mild assumptions where competitive exclusion holds, we show that:
\begin{itemize}
\item Each face of the PCH corresponds directly to a set of species that can coexist at steady state  (Figure \ref{fig:PCH}).
\item A face of the PCH has vertices $\mathbf C_{i_1},\dots,\mathbf C_{i_{S^\star}}$ if and only if there exists a choice of the resource supply vector $\mathbf K$ for which precisely the species $i=i_1,\dots,i_{S^\star}$ coexist. (Fig.~\ref{fig:PCH-3-resources})
\item Transitions between different steady-state behaviors as the resource supply vector $\mathbf K$ is varied (Figure \ref{fig:oldPicture}) are captured by the geometry of the PCH, with allowed transitions corresponding to neighboring faces (Fig.~\ref{fig:PCH-3-resources} and \ref{fig:facelattice}).
\end{itemize}

These three properties of the PCH allow us to enumerate, using only species attributes, possible ecologically stable steady states and transitions between them as the supplied resources are varied.
The information about steady-states and coexistence contained in the ZNGIs is exactly that contained in the PCH as they are dual objects.
The chief advantage provided by this dual picture is that it provides intuition for coexistence and niche differentiation in terms of species traits.
Additionally, the PCH construction lends itself well to computational geometry, especially in high dimensions as there are systematic and efficient algorithms for enumerating faces of a convex hull of a set of points \cite{barber1996quickhull}.
We now discuss the implications of this geometric picture in greater detail.

\subsection{Enumerating possible steady-states and transitions\label{sec:enuermating-and-transitions}}
\begin{figure*}[t!]
    \includegraphics[width=1.0\linewidth]{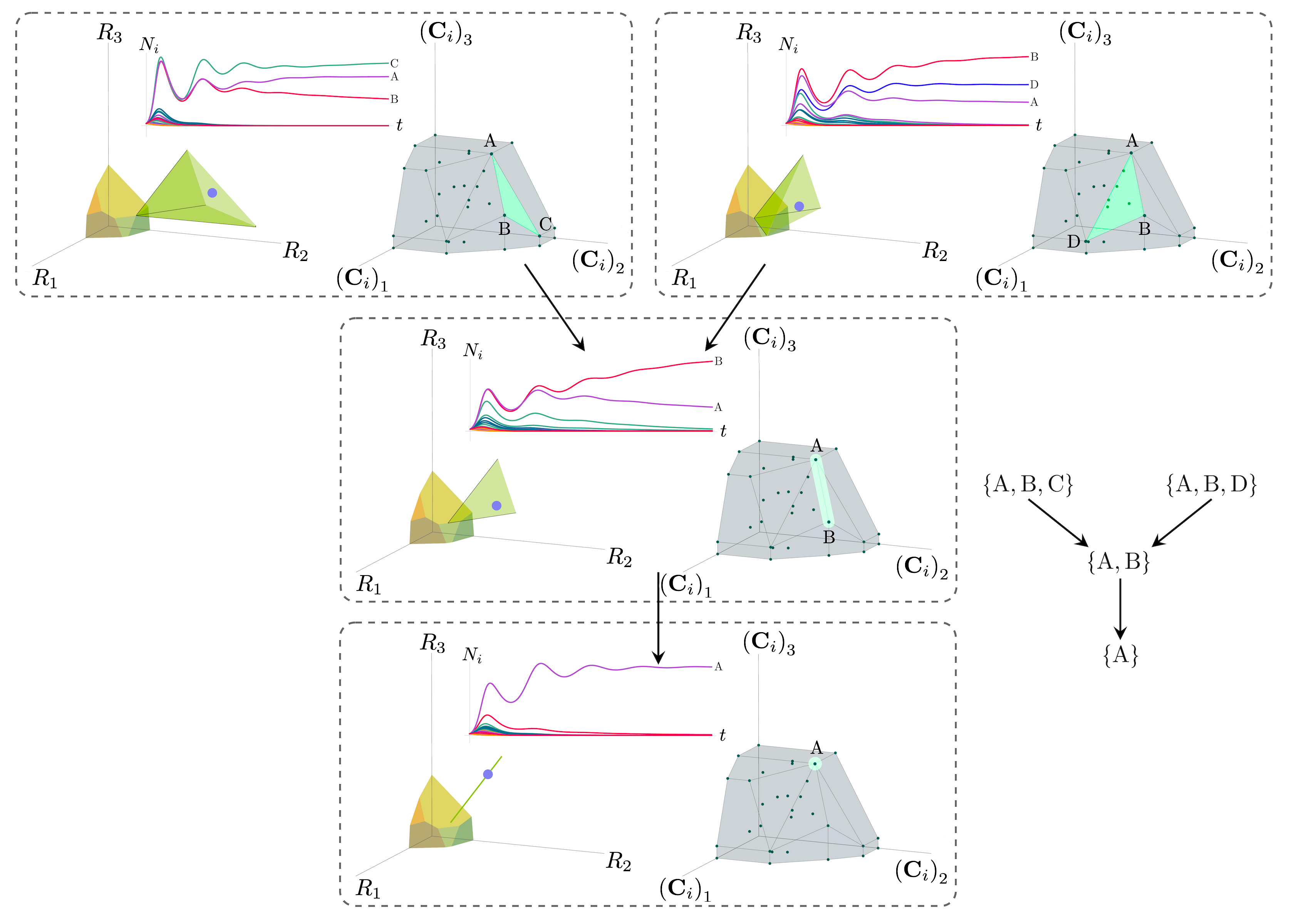}
    \caption[Outer hull with three resources]{
        Positive convex hull (PCH) formed by rescaled consumption vectors for an ecosystem with $M=3$ resources and $S=20$ species. In each panel, the left plot shows the infeasible region polytope formed by the ZNGI, the resource supply vector $\mathbf K$, and the associated coexistence cone.  The top plot shows the species populations $N_i$ as a function of time. The right plot depicts the PCH; the face corresponding to the steady-state for the choice of $\mathbf K$ is indicated by large green dots. Small dots indicate consumption vectors, $\mathbf{C}_i$, that lie in the interior of the PCH and correspond to species that go extinct for all choices of $\mathbf{K}$. The directed graph shows possible transitions between these steady-states as $\mathbf{K}$ is varied when no reinvasion occurs.
        A value of $\mathbf K$ is chosen arbitrarily in each pane such the species which survive demonstrate each face of the PCH. The panes are organized to reflect the face lattice of the PCH. (Also see the interactive {\em Mathematica} notebook.)
        }
        \label{fig:PCH-3-resources}
    \end{figure*}

Each face of the PCH is in direct correspondence to a possible set of species that can coexist. This allows us to easily enumerate all possible ecologically stable steady states by listing all faces of the PCH and their neighbors. A simple example of this can be seen by comparing Figures \ref{fig:oldPicture} and \ref{fig:PCH} which give two different geometric pictures for the same ecosystem with $S=5$ species and $M=2$ resources. Notice, there exists no choice of $\mathbf K$ where species $2$ and $3$ survive at steady-state. Both of these species are always competitively excluded from the ecosystem. In the PCH, this is reflected in the fact that the corresponding scaled consumption vectors fall on the interior of the PCH and hence are not part of any face. Furthermore, when $\mathbf{K}$ is varied in Fig.~\ref{fig:oldPicture}, the resulting transitions are precisely captured by the geometry of the PCH in Fig.~\ref{fig:PCH}: from a steady state where species $5$ survives, to coexistence of species 1 and 5, to a steady state with only species 1, to coexistence with species $1$ and $4$, and finally a steady state where only species 4 survives.
Here, we use the term ``transition'' to refer to a change in the set of species that coexist at uninvadable steady state as the resource supply vector $\mathbf K$ is varied either continuously or discretely while species may re-invade continuously.
These possible transitions are independent of how $\mathbf K$ is varied because for any choice of $\mathbf K$, there is a unique uninvadable steady state independent of initial conditions \cite{marsland2019minimum}.
If reinvasion did not occur, then the geometry of the PCH would change such that scaled consumption vectors for extinct species are removed from the construction; nonetheless, possibly coexisting communities would still be represented by faces of the PCH.

\begin{figure}[t]
    \centering
   \includegraphics[width=\linewidth]{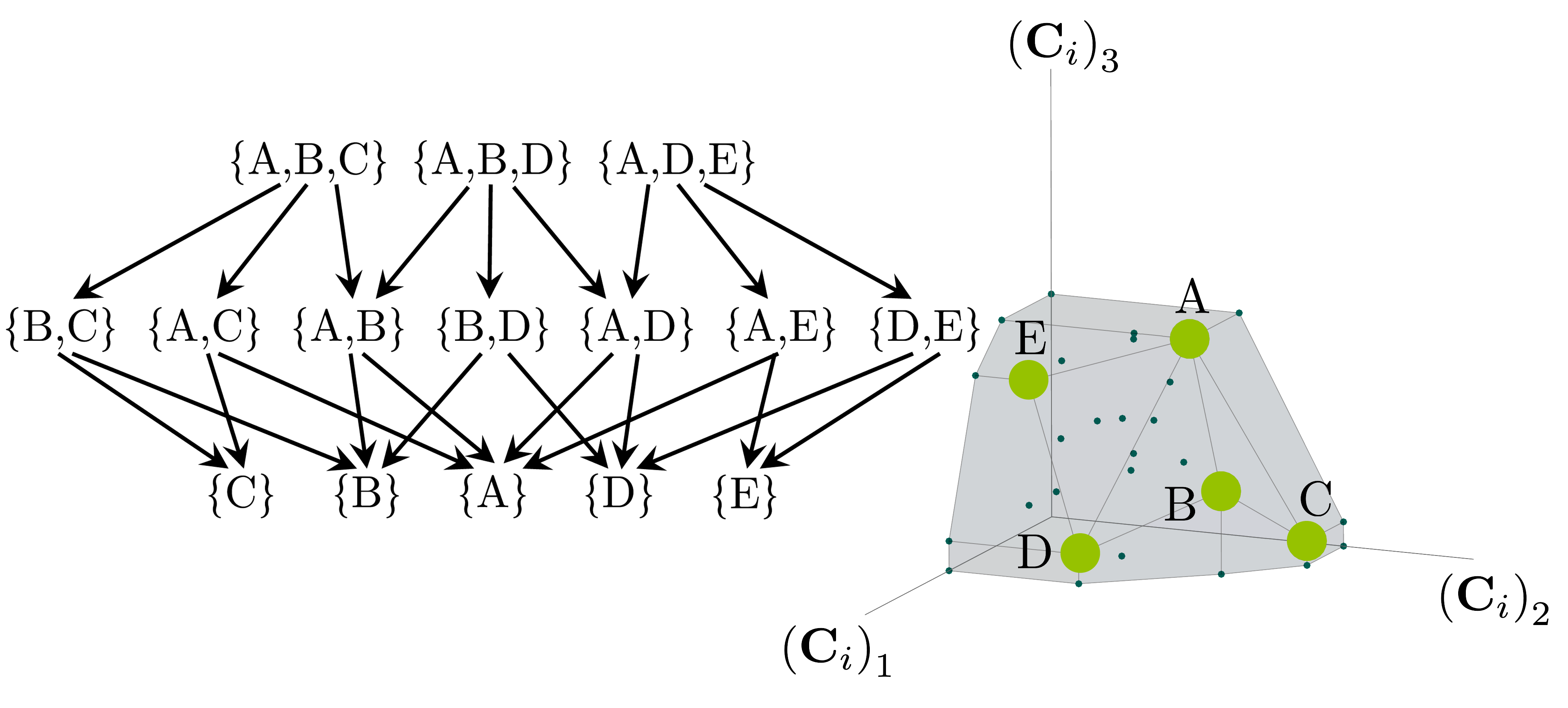}
    \caption{Enumerating steady-states and transitions using the PCH. All possible steady-states and allowed transitions (without reinvasion) for the ecosystem in Fig.~\ref{fig:PCH-3-resources} can be enumerated by considering the geometry of the faces on the PCH.}
        \label{fig:facelattice}
    \end{figure}

More generally, the faces of the PCH are arranged in a lattice by subset inclusion, so we can find all possible transitions between coexisting species by descending through the face lattice and enumerating the species whose scaled consumption vectors are the vertices of the faces. An example illustrating these transitions in a more complex ecosystem with $M=3$ resources and $S=20$ species is shown in Fig.~\ref{fig:PCH-3-resources} and the corresponding interactive {\em Mathematica} notebooks. The left side of each panel in the figure shows the geometry in resource abundance space and the location of the supplied resource vector $\mathbf{K}$, while the right-hand side shows the dual geometry in terms of the PCH. Notice that the faces of the PCH capture possible steady-states, and transitions between different steady states can only occur if faces are neighbors; the lattice of all faces and transitions is shown in Fig.~\ref{fig:facelattice}.

\subsection{Coexistence and competition}

\begin{figure*}[t!]
    \centering
    \includegraphics[width=0.8\linewidth]{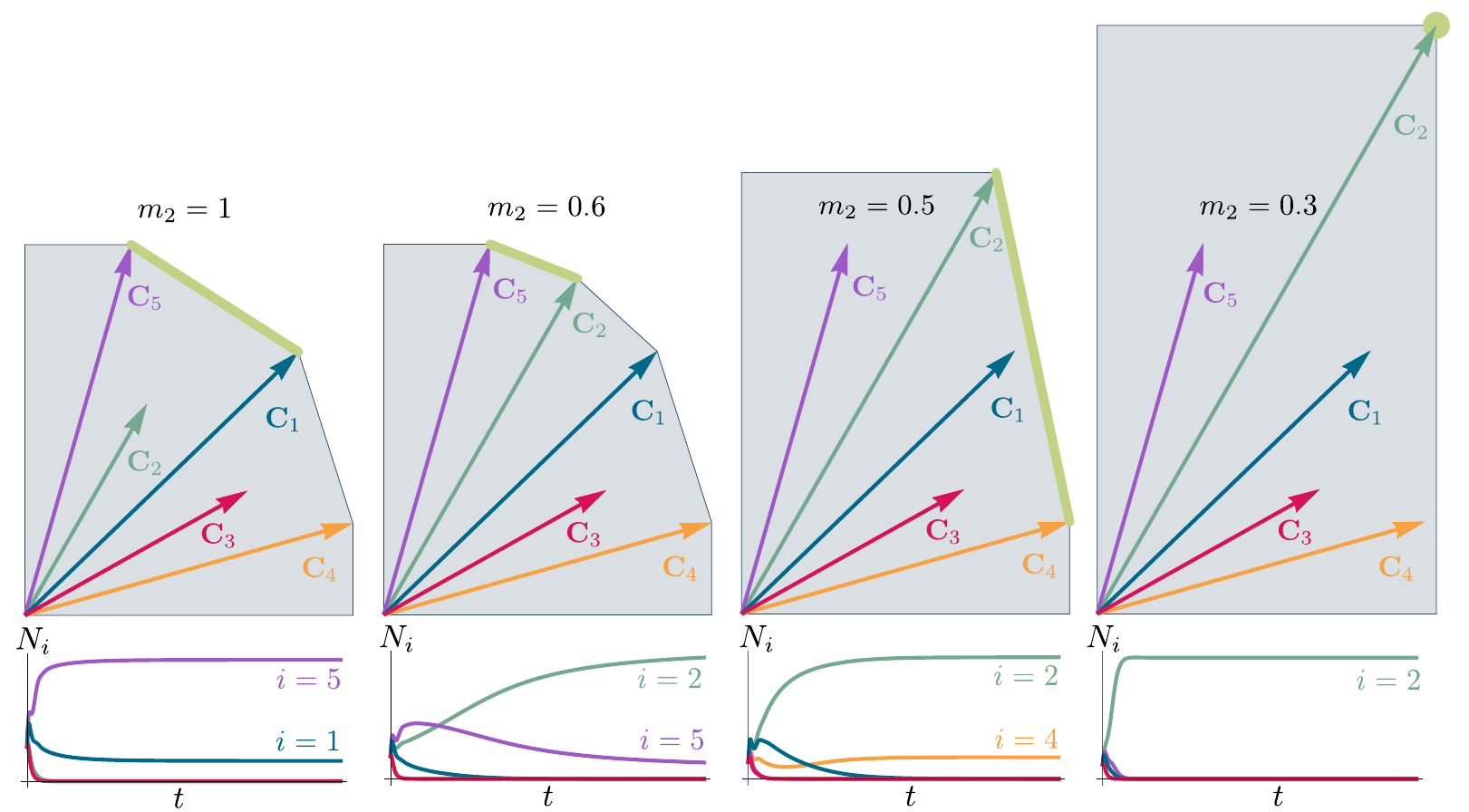}
    \caption[
        Modifying model parameters
    ]{
        Positive convex hull (PCH) of rescaled consumer preferences predicts coexistence properties when species fitness is modified. Panels show PCH (top) and species abundances as a function of time for fixed resource supply vector $\mathbf{K}$ (bottom) as the fitness of species $2$ is increased by decreasing the parameter $m_2$ for the ecosystem analyzed in Figures \ref{fig:oldPicture} and \ref{fig:PCH}.  Increasing the fitness of species 2 interrupts the coexistence of species 1 and 5, followed by different levels of competitive exclusion.
    }
    \label{fig:decreasing-death-rate}
\end{figure*}

The geometry of the PCH can also be used to ask and answer interesting ecological questions. For example, how does changing species traits such as consumer preferences, $c_{i \alpha}$, or fitnesses, $m_i$, affect what species can coexist? How does adding a fitter invasive species affect biodiversity?

To illustrate this, we once again consider the simple ecosystem of $S=5$ species and $M=2$ resources considered in Figures~\ref{fig:oldPicture} and \ref{fig:PCH} and ask how increasing the fitness (i.e., decreasing the death rate $m_2$) of species $2$ changes possible ecological steady-states.  Figure \ref{fig:decreasing-death-rate} shows the PCH and resource dynamics for this system as the fitness of species $2$ is increased. 

Notice, initially the rescaled consumption vector for species $2$, $\mathbf{C}_2$, lies within the PCH indicating that species $2$ always goes extinct for all choices of the resource supply vector $\mathbf{K}$. Furthermore, since species $1$ and $5$ lie on a common face of the PCH, we know that there exists a choice of $\mathbf{K}$ where species $1$ and $5$ coexist. The bottom panel shows the species abundances as a function of time for one such choice of $\mathbf{K}$ .

As we increase the fitness of species $2$ by decreasing $m_2$, the scaled species consumption vector $\mathbf{C}_2=w_2 \mathbf{c}_2 /m_2$ grows larger. Notice that the new PCH has faces with vertices of species $2$ and $5$ and species $2$ and $1$, but no face corresponding to species $1$ and $5$ as in the original PCH.
For this reason, species $1$ and $5$ can no longer coexist.  The simulations of species abundance as a function of time confirm this prediction. Further, increasing the fitness of species $2$ results in the competitive exclusion of species $1$ and $5$ and coexistence of species $2$ and $4$. Finally, when species $2$ is fit enough, it excludes all other species in the ecosystem.

This simple example helps illustrate how the PCH can be used to gain an intuitive understanding of how introducing a fitter species can change ecological steady states and biodiversity. If the fitness of the invasive species is the same order of magnitude as those of existing species, then its introduction will not result in large scale extinction so long as it is sufficiently distinct from existing species and does not have some special advantage in its consumption preferences.
Instead, the species will be able to carve out a distinct niche. However, as the magnitude of the rescaled consumer preference vector grows, there will be massive extinctions. The transitions between these two regimes can be encoded simply in the geometry of the PCH.
Additionally, the relative orientation of faces on the PCH can be used to understand how similar communities are to each other and the degree of competition between species.
Faces that are nearly parallel to each other indicate that the corresponding species are similar and will compete strongly and can survive in similar environments.
While these results can also be derived without the PCH by using the infeasible region, the PCH provides many intuitions directly in terms of species' attributes.
Finally, while here we have restricted our considerations to a low-dimensional setting with two resources, we expect new and interesting properties to emerge in high-dimensional settings with many resources since high-dimensional convex geometry is much more complex than its low-dimensional analog.

\subsection{Alternative consumer resource models\label{sec:other-models}}

In the previous section, we restricted our analysis to MacArthur's original consumer resource model. However, the basic geometric picture discussed above also holds for other popular variants of consumer resource models including the Tilman's consumer resource model (TCRM),
\begin{align}
    \frac{\mathrm dR_\alpha}{\mathrm dt}&= \left(K_\alpha - R_\alpha
    \right)-\sum_{i=1}^S c_{i\alpha} N_i,  \nonumber \\
   \frac{\mathrm dN_i}{\mathrm dt} &=N_i \left(\sum_{\alpha = 1}^M w_\alpha c_{i\alpha} R_\alpha-m_i \right),
     \label{Eq:populationTilman}
\end{align}
and the consumer resource model with externally supplied resources (eCRM, Eq.~\eqref{Eq:populationExt}),
\begin{align}
    \frac{\mathrm dR_\alpha}{\mathrm dt}&= \left(\kappa_\alpha - R_\alpha
    \right)-\sum_{i=1}^S c_{i\alpha} N_i R_\alpha, \nonumber \\
   \frac{\mathrm dN_i}{\mathrm dt} &=N_i \left(\sum_{\alpha = 1}^M w_\alpha c_{i\alpha} R_\alpha-m_i \right).
     \label{Eq:populationExt}
\end{align}
The underlying reason for this is that our geometric construction is derived from analyzing the sign of $g_i(\mathbf R^\star)$ for each species and the form of $g_i$ is identical for the eCRM, TCRM, and MCRM. The faces of the convex polytope formed from the rescaled consumption vectors still enumerate subsets of species that can stably coexist in each of these models. For the eCRM, this construction is nearly identical to the MCRM; while the positions of the coexistence cones are different, their structure and the PCH of rescaled consumption vectors are identical. The correspondence between stably coexisting species in the eCRM and faces of the PCH is shown in Figs.~\ref{fig:eCRM-PCH} and \ref{fig:eCRM-3D}. For the TCRM, the appropriate convex polytope constructed from rescaled consumption vectors must be slightly modified because resource abundances can become nonphysically negative; for discussion and visualization, see appendix \ref{appendix:TCRM}.

\subsection{Extension to models with nonlinear species growth-rates}

The geometric construction presented in the last section exploits that fact that the growth rate of a species $g_i(\mathbf{R})$ is a linear function of the resource abundances. Here, we show that many of the geometric intuitions still hold even in more complex models where $g_i(\mathbf{R})$ is a nonlinear function. To do so we consider a general consumer resource model of the form,
\begin{align}
  \frac{\mathrm dN_i}{\mathrm dt}
  &=N_i g_i(\mathbf R),\\
  \frac{\mathrm dR_\alpha}{\mathrm dt}
  &=h_\alpha(\mathbf R) - \sum_i N_i q_{i\alpha}(\mathbf R),
 \label{Eq:NLCRM}
\end{align}
where the  $q_{i\alpha}(\mathbf R)$ are impact vectors that encode how species change resource abundances. We further restrict our considerations to the case where,
\begin{align}
  q_{i\alpha}(\mathbf R)
  =a_i(\mathbf R) b_\alpha(\mathbf R) \frac{\partial g_i}{\partial R_\alpha},
  \label{Eq:MEPP-sym-condition}
\end{align}
holds, with $a_i(\mathbf R)$ and $b_\alpha(\mathbf R)$ arbitrary functions. Such a form can be motivated by noting that this choice reflects the idealized case where species consume resources directly proportional to the marginal utility they derive. It was also shown in Ref.~\cite{marsland2019minimum} that for such a choice, interactions between species are always symmetric and there is a unique invadable fixed point.
The uniqueness of an invadable fixed point when this condition holds is important because it implies that the PCH is well-defined and transitions between different steady states are invariant to the initial conditions, as discussed in Sec.~\ref{sec:enuermating-and-transitions}.

\begin{figure}
  \includegraphics[width=\linewidth]{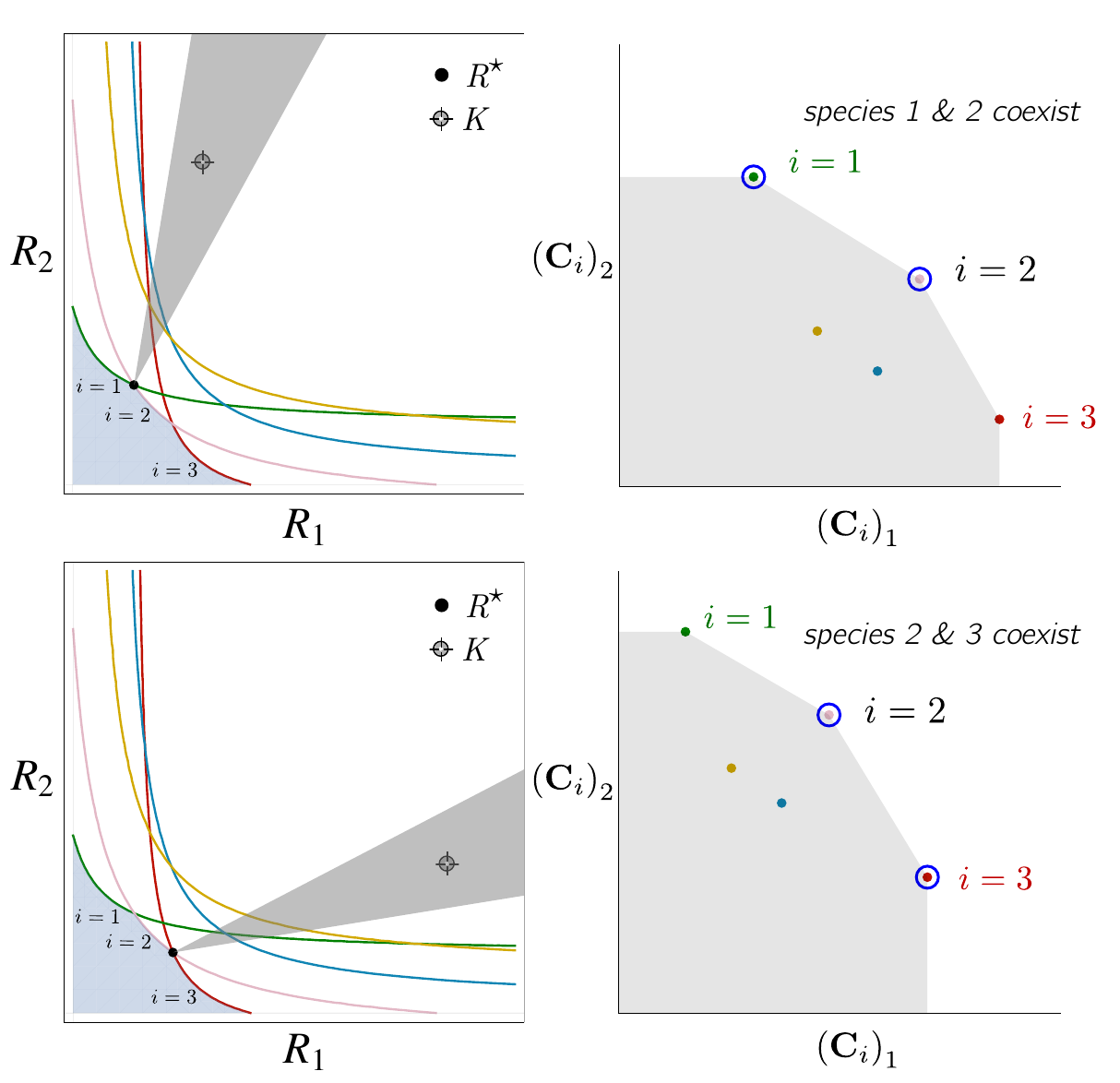}
  \caption{
    \label{fig:NonlinearPCH}
    Positive convex hull (PCH) construction for a CRM with a nonlinear growth rate.
    Each row corresponds to a value of $\mathbf R^\star$; the left panels show $\mathbf K, \mathbf R^\star$, ZNGIs, coexistence cones; the right panels show the PCH construction and surviving species for the given $\mathbf R^\star$.
    For a nonlinear growth rate, $g_i(\mathbf R)$, the PCH's form depends on $\mathbf R^\star$.
    This figure corresponds to the model from Eqs.~\eqref{eq:nonlinExampleModelR} and \eqref{eq:nonlinExampleModelg}.
  }
\end{figure}

For this more general nonlinear case, many properties of the relationship between the PCH and coexistence are preserved, but unlike in the linear case, the consumption vectors $\mathbf C_i(\mathbf R^\star)$ now depend on the steady-state resource abundances, $\mathbf R^\star$.
Expanding the species' growth rates about $\mathbf R = \mathbf R^\star$ to linear order gives:
\begin{align}
  g_i(\mathbf R)
  \approx
  \sum_{\beta = 1}^M
  \left.\frac{\partial g_i}{\partial R_\beta}\right|_{\mathbf R = \mathbf R^\star} (R_\beta - R_\beta^\star)
  +
  g_i(\mathbf R^\star).
\end{align}
Defining the $\alpha$th component of $\mathbf C_i(\mathbf R^\star)$ to be,
\begin{align}
  [\mathbf C_i (\mathbf R^\star)]_\alpha
  =
  \frac{
  \left.\frac{\partial g_i}{\partial R_\alpha} 
    \right|_{\mathbf R = \mathbf R^\star}}{
      \sum_{\beta = 1}^M 
      \left.\frac{\partial g_i}{\partial R_\beta} 
      \right|_{\mathbf R = \mathbf R^\star}
      R_\beta^\star
      -
      g_i(\mathbf R^\star)
    },
\end{align}
to linear order $g_i(\mathbf R^\star) \leq 0 $ is equivalent to,
\begin{align}
  \mathbf C_i (\mathbf R^\star) \cdot \mathbf R^\star - 1 \leq 0.
\end{align}
As shown in Fig.~\ref{fig:NonlinearPCH}, the consumption vectors now depend on the steady-state concentration $\mathbf R^\star$, and both the PCH and its faces deform as $\mathbf R^\star$ changes as parameters are varied. However, the PCH still contains a considerable amount of information about possible steady-states and transitions. Given steady-state resource and species abundances $\mathbf R^\star$ and $N_i^\star$, the consumption vectors corresponding to the surviving species still span a face of the PCH. In addition, we have found the PCH generally also captures the local steady-state structure including local transitions in steady-state behavior as a function of model parameter. The reason for this is that the structure of the PCH varies smoothly with $\mathbf R^\star$, so transitions between stably coexisting subsets of species are represented by adjacent faces on the PCH  (see the interactive demonstrations (appendix \ref{MMA-notebook-links}) to further understand utility and limitations). 

One crucial difference is that in the nonlinear case not all faces present on a specific realization of the PCH (i.e. a PCH corresponding to a particular choice of parameters) necessarily represent realizable steady-states. This is because as $\mathbf R^\star$ and $N_i^\star$ change, new faces can form and disappear. Since the PCH depends on linearizing the growth rate, it does not have any global information about nonlinear effects.
Information about these nonlinear effects is captured fully by looking to the infeasible region as described in Sec.~\ref{sec:resource-space-geom}.
However, with a moderate number of resources, the infeasible region in high-dimensions is difficult to visualize and compute, especially for nonlinear growth rates.
Because the PCH can be computed efficiently using convex hull algorithms, it is a useful tool for understanding the local structure of transitions between steady-states in high-dimensional systems, even when growth rates may be nonlinear.

In Fig.~\ref{fig:NonlinearPCH}, we demonstrate how the PCH extends to the nonlinear case by considering a consumer resource model where species growth saturates as a function of resource concentrations (also called a Type II functional response in the ecological literature),
\begin{align}
  \frac{\mathrm dR_\alpha}{\mathrm  dt}
  &=
  \tau^{-1} (K_\alpha - R_\alpha)
  -
  \sum_{i=1}^S N_i R_\alpha \frac{e_i k_{i\alpha} \mu_{i\alpha}}{(k_{i\alpha} + R_\alpha)^2},
  \label{eq:nonlinExampleModelR}
  \\
  g_i(\mathbf R)
  &=
  e_i \sum_{\alpha=1}^M \frac{\mu_{i\alpha} R_\alpha}{k_{i\alpha} + R_\alpha} - m_i,
  \label{eq:nonlinExampleModelg}
\end{align}
with $M=2$ resources and $S = 5 $ species \footnote{~
The form of the species growth and resource impact rates differ (the denominators are linear and quadratic, respectively) so that the appropriate symmetry condition Eq.~\ref{Eq:MEPP-sym-condition} is satisfied.~
}.
In the figure, one sees that as $\mathbf K$ changes, the resulting steady-state resource abundances $\mathbf R^\star$ also change, as do the generalized consumption vectors $\mathbf C_i(\mathbf R^*)$ and the resulting PCH. Since $\mathbf R^\star$ depends smoothly on the parameters,  both $\mathbf C_i(\mathbf R^*)$ and the PCH also smoothly change as parameters are varied. For this reason, at least locally, the PCH remains an important source of intuition for possible ecological behaviors.

\subsection{Limitations of geometric construction}

We end by briefly commenting on some important limitations of our geometric construction. 
In general, we expect that our geometric construction is valid whenever we can use ideas from contemporary niche theory like ZGNIs and coexistence cones to describe the underlying ecology \cite{chase2009ecological, tilman1982resource, levins2020evolution}. The underlying reason for this is that the geometry in the space of consumer preferences is mathematically dual to the geometry in resource abundance space. The validity of both geometric  constructions require that the consumption vectors (i.e., how resources affect growth rate) and the impact vectors (i.e., how species deplete resources) be closely related to each other. This statement can be made more precise using the recently discovered relationship between consumer resource models and constrained optimization \cite{mehta2019constrained, marsland2019minimum}. We expect the geometric pictures to hold when such a mapping to constrained optimization exists, namely that the effective species-species interactions are symmetric \cite{marsland2019minimum}. 

Another limitation of our construction is that it assumes that there are no hard geometric constraints on consumer-resource preferences. One prominent example of such hard constraints are metabolic tradeoffs that fix the consumer preference vectors to all have the same magnitude \cite{posfai2017metabolic}. Such constraints result in the violation of competitive exclusion and require evolutionary fine-tuning of preferences \cite{d2020emergent,pacciani2020dynamic}. In this limit, our geometric picture begins to break down because the very tightly spaced faces become essentially parallel and species belonging to multiple faces can coexist (see Appendix~\ref{justification}). In essence, we can no longer think of our geometry as a convex shape but instead must treat it as a smooth sphere. 
As soon as some constraints are relaxed very slightly---perhaps by some disordered perturbation---the geometric principles become valid again with a very large number of faces which are very small, tightly packed, and nearly parallel.
This indicates that the species present in a community are very sensitive to the supply point and transitions between communities happen readily.

\section{Discussion}
    
In this work,  we have introduced a new geometric framework for understanding niche theory based on the species consumer resource preferences. Our work complements existing geometric intuitions by emphasizing the important role played by consumer preferences in shaping species coexistence and niche differentiation. One appealing aspect of the work is that it works in trait space, something that is often easier to observe and measure than resource abundances \cite{mcgill2006rebuilding}. Despite the simplicity of our picture, it can be used to make a series of powerful predictions including which species can coexist and how species coexistence patterns can transition as the resource supply vector is varied. 
We hope that our geometric framework can be used to help design new experiments and analyze empirical directions.


  The geometric picture developed here may extend to other contexts to provide new insights. In this manuscript, we have largely focused on small ecosystems with a few resources and species. It will be interesting to ask how this picture generalizes to large ecosystems where methods from random matrix theory and statistical physics can be used to make powerful predictions \cite{allesina2015stability}. Doing so will require understanding convex hulls of randomly distributed points and represents an interesting mathematical problem \cite{litvak2005smallest}. Another direction to explore is how this analysis can be extended to consider ecosystems with multiple trophic layers \cite{duffy2007functional} and to microbial ecosystems where metabolic cross-feeding plays a central role \cite{goldford2018emergent, amarnath2021stress, dal2021resource}. Additionally, we may be able to extend this picture to understand temporal niches, which may play an important role in shaping microbial ecosystems \cite{amarnath2021stress, fridman2022fine}.

\section{Acknowledgements} We thank Zhijie (Sarah) Feng and Jason Rocks for useful discussions. We also thank Emily Hager and Maria Yampolskya for comments on the manuscript. This work was funded by NIH NIGMS Grant No. R35GM119461 to P.M. and the Boston University Undergraduate Research Opportunities Program to E.B.

\appendix

\section{Details of models analyzed in figures\label{appendix:ModelsInFigures}}

\subsection{Two resource model\label{appendix:TwoResourceModel}}

In all figures with two resources $(M=2)$, the MCRM differential equations are simulated with five species $(S=5)$ and the following parameters:
\begin{gather}
  [c_{i\alpha}]
  =
  \left[
\begin{array}{cc}
 0.7185 & 0.202 \\
 0.2675 & 0.464 \\
 0.2325 & 0.812 \\
 0.599 & 0.577 \\
 0.4875 & 0.273 \\
\end{array}
\right]
\\
m_i = 1, \quad i =1 ,\dots, 5
\\
w_\alpha = 1, \quad \alpha = 1,2.
\end{gather}
The values of $c_{i\alpha}$ were chosen by-hand to make features of the PCH clear.
The various choices of $K_\alpha$ were also chosen by hand to highlight various coexistence cones; in Fig.~\ref{fig:oldPicture}, $\mathbf K_A = (0.53,2.48), \mathbf K_B = (1.48,2.82), \mathbf K_C = (1.92,1.74), \mathbf K_D = (2.88,1.18), \mathbf K_E = (2.85,0.50)$; in Fig.~\ref{fig:decreasing-death-rate}, $\mathbf K = (1.21, 2.82)$.
These values can be modified by-hand using a click-and-drag interface in the interactive {\em Mathematica} demonstration.
In the numerical solutions to the differential equations, the initial conditions used are: $N_i(t=0) = 0.5$ and $R_\alpha(t=0) = K_\alpha$.
In plots of the PCH and consumption vectors, the horizontal and vertical axes range from 0 to 1; in plots of coexistence cones and the infeasible region, the horizontal and vertical axes range from 0 to 3.
The vertical axes of plots of species population dynamics are scaled so that the curve fills the entire plot; simulations are run and plotted until $t=100$.
For further details on the plots and numerical solutions including code or to modify plotting parameters, see the interactive demonstration.

\subsection{Three resource model}

\begin{figure*}[t]
  \includegraphics[width=1.0\linewidth]{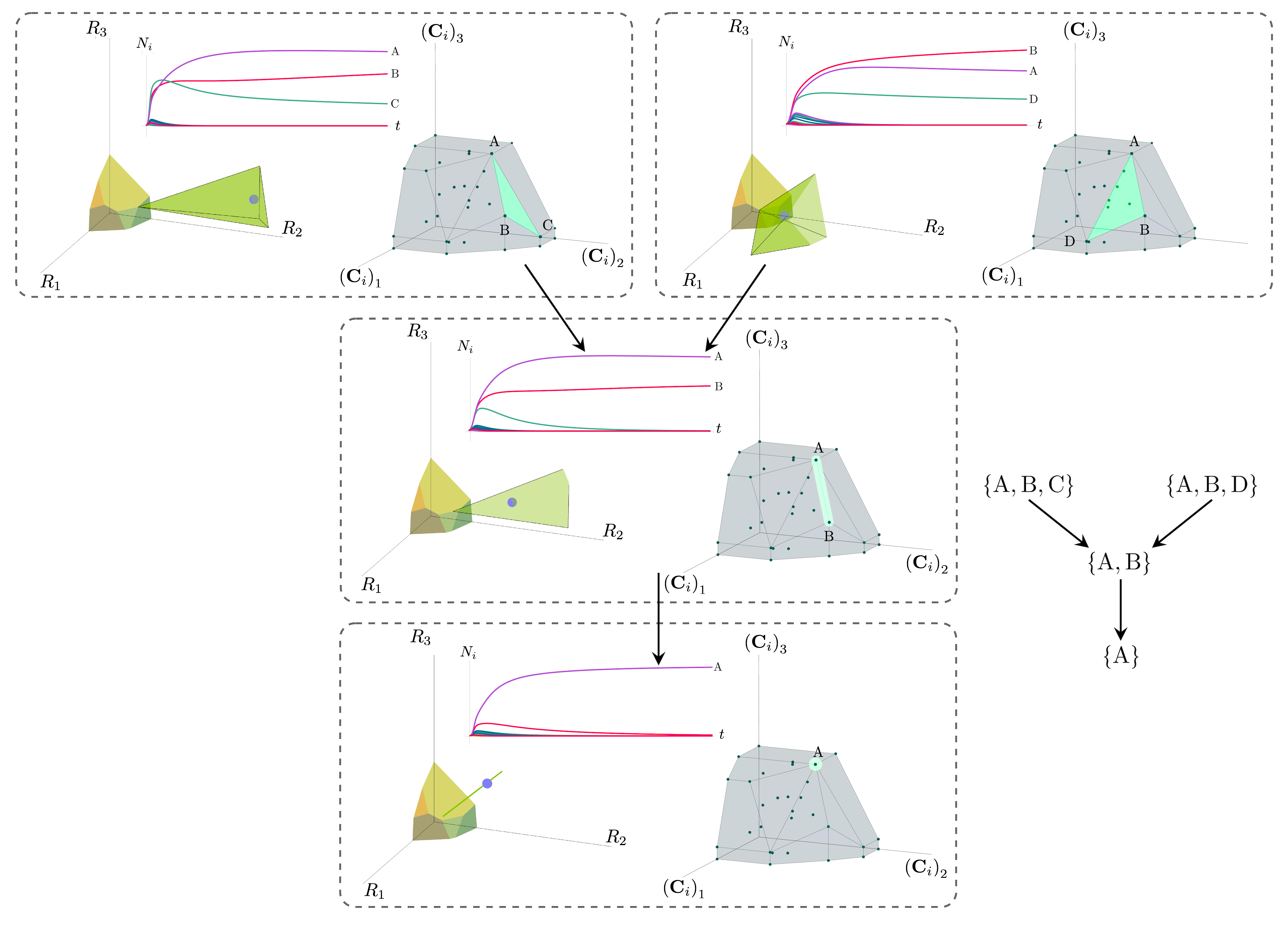}
  \caption{\label{fig:eCRM-3D}
    Positive convex hull (PCH) formed by rescaled consumption vectors for an ecosystem with $M=3$ resources with dynamics given by the consumer resource model with externally supplied resources (eCRM).
    The structure and layout of this figure is identical to that of Fig.~\ref{fig:PCH-3-resources}, but the dynamics and coexistence cones shown are those of the eCRM model (Eq.~\ref{Eq:populationExt}).
  }
\end{figure*}

In all figures with three resources $(M=3)$ the MCRM differential equations are simulated with twenty $(S=20)$ species and the following parameters:
\begin{gather}
  [c_{i\alpha}]
  =
  \left[
\begin{array}{ccc}
 0.416 & 0.023 & 0.17 \\
 1.652 & 0.923 & 0.198 \\
 0.329 & 0.874 & 0.8 \\
 0.135 & 0.565 & 0.741 \\
 0.226 & 0.173 & 1.109 \\
 0.069 & 0.616 & 1.262 \\
 0.424 & 0.777 & 1.394 \\
 0.17 & 0.401 & 0.716 \\
 1.074 & 1.041 & 0.643 \\
 1.038 & 1.666 & 0.555 \\
 0.184 & 0.926 & 1.029 \\
 0.588 & 2.077 & 0.163 \\
 1.466 & 0.881 & 0.144 \\
 1.221 & 0.38 & 1.211 \\
 0.888 & 0.891 & 0.016 \\
 0.024 & 0.077 & 0.535 \\
 1.131 & 0.534 & 0.465 \\
 0.797 & 1.337 & 1.49 \\
 0.914 & 0.988 & 0.716 \\
 0.295 & 1.071 & 0.531 \\
\end{array}
\right],
\\
m_i = 1,\quad i=1,\dots, 20,
\\
w_\alpha = 1, \quad \alpha = 1,2,3.
\end{gather}
The values of $c_{i\alpha}$ were drawn as random variables $|X|$, where $X$ is a unit normal random variable.
The various choices of $K_\alpha$ were chosen by-hand to highlight various coexistence cones; in Fig.~\ref{fig:PCH-3-resources}, upper left: $\mathbf K = (0.64,1.49,0.64)$, upper right: $\mathbf K = (0.94,1.00,0.55)$, center: $\mathbf K = (0.81,1.04,0.57)$, bottom: $\mathbf K = (0.66,0.84,0.92)$; in Fig.~\ref{fig:eCRM-3D}, upper left: $\pmb \kappa = (0.58,1.87,0.53)$, upper right: $\pmb \kappa = (1.41,1.02,0.53)$, center: $\pmb \kappa = (0.94,1.21,0.69)$, bottom: $\pmb \kappa = (1.10,1.13,1.01)$.
These values can be modified using a drag-and-drop interface in the interactive {\em Mathematica} demonstrations.
In the numerical solutions, the initial conditions used are $N_i(t=0) = 0.01$ and $R_\alpha = K_\alpha$.
In plots of the PCH and consumption vectors, the all axes range from 0 to 3; in plots of coexistence cones and the infeasible region, the horizontal and vertical axes range from 0 to 2.
The consumption vectors that are labeled A, B, C, and D correspond to species $i = 18,i=10,i=12,i=2$ in the above matrix, respectively.
The vertical axes of plots of species population dynamics are scaled so that the curve fills the entire plot; simulations are plotted until $t=40$.

\subsection{Model with nonlinear growth rate\label{appendix:NonLinFigureDetails}}

Figure \ref{fig:NonlinearPCH} is constructed using the CRM,
\begin{align}
  \frac{\mathrm dR_\alpha}{\mathrm dt}& = \tau^{-1}(K_\alpha - R_\alpha) - \sum_{i=1}^S N_i R_\alpha \frac{e_i k_{i\alpha} \mu_{i\alpha}}{(k_{i\alpha} + R_\alpha)^2},
  \\
  \frac{\mathrm dN_i}{\mathrm dt} &= N_i g_i(\mathbf R),
\end{align}
which has nonlinear growth rate,
\begin{align}
  g_i(\mathbf R)
  =
  e_i\sum_{\alpha = 1}^M \frac{\mu_{i\alpha} R_\alpha}{k_{i\alpha} + R_\alpha} - m_i.
\end{align}
In the figure, there are $M =2$ resources and $S =5$ species with parameters,
\begin{gather}
  [\mu_{i\alpha}]
  =
\begin{bmatrix}
 2 & 0.5 \\
 0.5 & 2 \\
 1.5 & 1.5 \\
 0.9 & 0.7 \\
 0.7 & 0.9 \\
\end{bmatrix},
\quad
[k_{i\alpha}]
=
\begin{bmatrix}
 2.03 & 0.54 \\
 0.54 & 2.05 \\
 1.43 & 1.58 \\
 0.96 & 0.77 \\
 0.63 & 0.81 \\
\end{bmatrix},
\\
[m_i]
=
\begin{bmatrix}
  1 & 1 & 1.1 & 1 & 1 
\end{bmatrix},
\quad \tau = 1,
\\
e_i = 1,\quad i =1,\dots,5.
\end{gather}

\subsection{Interactive demonstrations\label{MMA-notebook-links}}

Interactive {\em Mathematica} visualizations are available in notebook-form at the following link:
\url{https://github.com/Emergent-Behaviors-in-Biology/Geometry-of-Coexistence}

\section{Technical details of geometric construction\label{justification}}
\begin{figure}[t]
  \includegraphics[width=\linewidth]{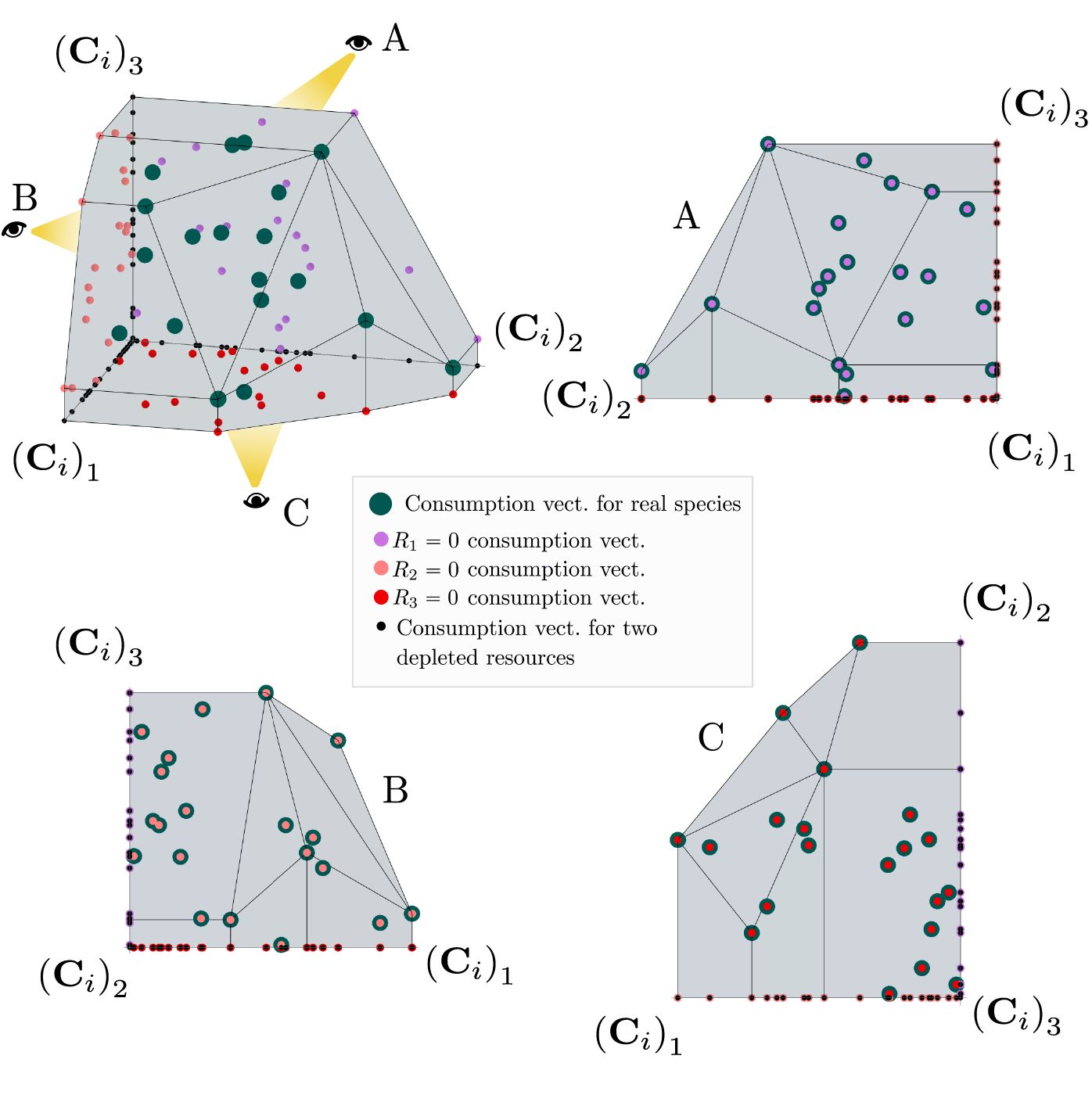}
 \caption[Resource depletion conditions]{
   Construction of the positive convex hull of consumption vectors with three resources.
   Upper left: the positive convex hull of rescaled consumption vectors, represented by large blue dots.
   The projection of the rescaled consumption vectors onto the coordinate planes and axes are displayed with differently colored smaller dots.
   Different view perspectives labeled A, B, C are shown with small eye cartoons.
   Upper right, lower left, lower right:
   The PCH along with the projected rescaled consumption vectors are shown from the different labeled view perspectives, emphasizing that the construction of the PCH is preserved under projection onto coordinate planes.
   }
 \label{fig:PCH-project}
\end{figure}
\begin{figure}[t]
  \includegraphics[width=\linewidth]{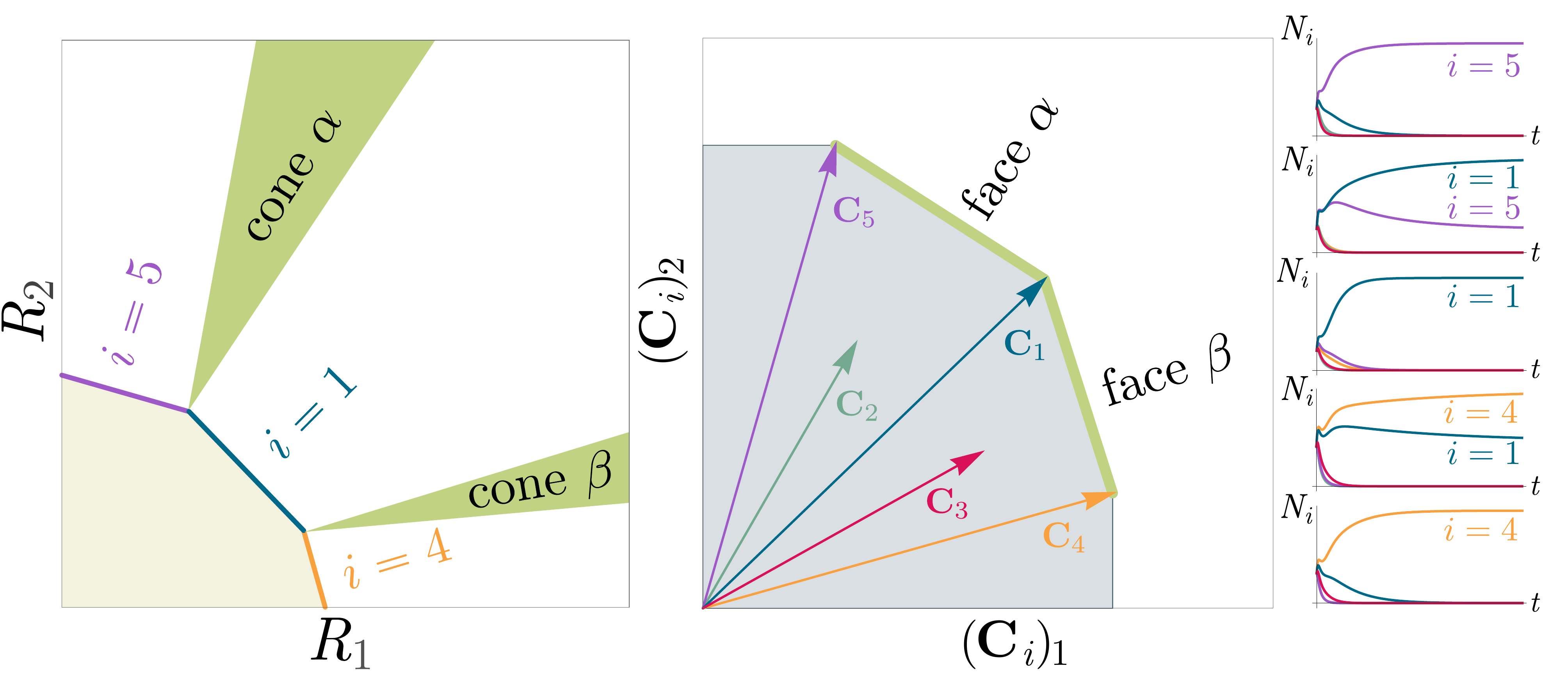}
   \caption{
    Relationship between coexistence cones, the PCH formed by rescaled consumption vectors, and stably coexisting species in the consumer resource model with Externally supplied resources (eCRM).
    Left: Coexistence cones and ZNGIs are highlighted in the resource-abundance phase space.
    Notice that the coexistence cones have the same structure but are in different positions compared to fig. \ref{fig:PCH}
    Center: PCH formed by rescaled consumption vectors; the faces of this PCH enumerate stably coexisting species.
    (c.f., fig. \ref{fig:PCH}).
    Right: Population dynamics of all sets of stably coexisting species; each plot corresponds to a different choice of supply rates, $\kappa_\alpha$.
  }
  \label{fig:eCRM-PCH}
\end{figure} 
\subsection{Resource depletion/defining the positive convex hull\label{appendix:PCHappendix}}


Let $\{\mathbf C_i\}_{i=1}^n \in \R^M$ be a set of points with components $\mathbf C_i = \left(
    \begin{array}{ccccc}
        C_{i1}&
        \dots&
        C_{i\alpha}&
        \dots&
        C_{iM}
    \end{array}
\right)^{\mathrm{T}}
$ which are all nonnegative: $C_{i\alpha} \geq 0$.
If $A \subseteq \{1,\dots,M\}$ is a subset of resources, then let $P_A$ project a vector $\mathbf C_i$ onto the coordinate axes $\alpha \in A$; that is, $P_A(\mathbf C_i)$ has the entries of $\mathbf C_i$ but with $C_{i\alpha'} = 0$ for entries with indices $\alpha' \notin A$.
The positive convex hull (PCH) is the convex hull of $\cup_{i \in \{1,\dots,S\}}\cup_{A \in 2^{\{1,\dots,M\}}} P_A(\mathbf C_i)$, the union of $P_A(\mathbf C_i)$ over all possible subsets $A$ of $\{1,\dots,M\}$ and all species $i \in \{1,\dots,S\}$.
The PCH is the convex hull of a total of $S \times 2^M$ points in $\R^M$, including the origin.
The PCH occurs in this work because the concepts presented should be consistent when any subset of resources is removed/depleted from the ecosystem.
For an example of this consistency under projection, see Fig.~\ref{fig:PCH-project}.
The additional points created by projection $P_A(\mathbf C_i)$ could be interpreted to represent specialist species that are consumed only by a subset, $A$, of resources.
The addition of these specialist species allows Eq.~\eqref{Eq:cone-eq} to hold even when resources are depleted.

\subsection{Deriving properties of the PCH\label{appendix:PCHproperties}}
We will show that the consumption vectors that form vertices of a common face of the positive convex hull correspond directly to a set of ecologically stable coexisting species.
Specifically, we will use the definition that if $\mathbf R^\star$ are the steady-state resource abundances then a species $i$ can survive if $\mathbf C_i \cdot \mathbf R^\star = 1$.

When we form the positive convex hull (PCH) of $\{\mathbf C_1,\dots,\mathbf C_S\}$, every point $\mathbf x$ in the positive orthant of $\R^M$ is contained in the conic hull of a face of the PCH.
If $\mathbf C_1,\dots, \mathbf C_k,\dots, \mathbf C_d$ are the vertices of the minimal face whose conic hull contains $\mathbf x$, then there are $\beta_k > 0$ so that $\mathbf x = \sum_{k =1}^d \beta_k \mathbf C_k$; note that $\mathbf C_k$ may include consumption vectors that have been projected onto a subset of axes, as in Appendix~\ref{appendix:PCHappendix}.
If $\mathbf x$ is in the interior of the PCH, then $\sum_{k=1}^d \beta_k < 1$; if $\mathbf x$ is on the boundary of the PCH, then $\sum_{k=1}^d \beta_k = 1$; and if $\mathbf x$ is outside the PCH, then $\sum_{k=1}^d \beta_k > 1$.

Let $j=1,\ldots, S^*$ index the surviving species, so $\mathbf C_j \cdot \mathbf R^\star = 1$ for all choices of $j$. If $\mathbf x = \sum_{j} \gamma_j \mathbf C_j$ with $\gamma_j \geq 0$ and $\sum_{j} \gamma_j = 1$ (i.e., $\mathbf x$ is a convex combination of $\mathbf C_j$), then $\left(\sum_{j} \gamma_j \mathbf C_j \right) \cdot \mathbf R^\star = \sum_{j} \gamma_j  = 1$.
Now, let $\mathbf x = \sum_{k=1}^d \beta_k \mathbf C_k$ where $\beta_k >0$ are the minimal face conic hull coordinates like before; because $\mathbf C_i \cdot \mathbf R^\star \leq 1$ for all species $i$, if $\mathbf x$ is in the interior of the PCH, then $\mathbf x\cdot \mathbf R^\star = \sum_{k} \beta_k \mathbf C_k \cdot \mathbf R^\star \leq \sum_{k} \beta_k < 1$.
Therefore, $\mathbf x \cdot \mathbf R^\star = 1$ implies that $\mathbf x$ is on the boundary or outside the PCH; convex combinations of points in a convex set cannot be outside the convex set, so $\mathbf x$ must be on the boundary of the PCH.
If all convex combinations of vertices of a convex polytope are on the boundary of the convex polytope, then the vertices belong to a common face of the convex polytope.

Conversely, if there is a face with vertices indexed by $j$, then we can find a choice of $\mathbf R^\star$ (although it may not be unique if the number of surviving species is less than $M$) so that $\mathbf C_j \cdot \mathbf R^\star = 1$ for all $j$ and $\mathbf C_k \cdot \mathbf R^\star < 1$ for $k \ne j$.
We can guarantee there is a choice of $\mathbf R^\star$ so that $\mathbf C_j \cdot \mathbf R^\star = 1$ because in $\R^M$, there is always an intersection of at most $M$ hyperplanes which are not parallel.
We then know that $\mathbf R^\star$ satisfies $\mathbf C_k \cdot \mathbf R^\star < 1$ for $k \ne j$ because, following similar logic from before, if $\mathbf C_k \cdot \mathbf R^\star \geq 1$, then we should be able to form any nontrivial convex combination of $\mathbf C_k$ and $\mathbf C_j$, and it will on the boundary or outside the PCH; however, no nontrivial convex combinations of vertices that do not belong to a common face can lie on the boundary of the PCH, so we have a contradiction.
The carrying-capacities $\mathbf K = \mathbf R^\star + \sum_{j}  \mathbf c_j $ will then lead to all species $j$ surviving.

In this argument, we assumed that there are no $\mathbf C_i$ that lie on a face of which it is not vertex.
Equivalently, we assumed that there are no faces that contain a subface of the same rank.
We could also say that there are no parallel faces.
If this assumption is dropped, then the statement `all convex combinations of vertices of a common face belong to just one face' no longer holds; some convex combinations could belong to any of the various subfaces.
This means that all $\mathbf C_i$ that belong to the face which contains all other subfaces---whether the $\mathbf C_i$ are vertices are not---represent species that can coexist.
In cases like these, it is possible that the number of surviving species is greater than the number of resources.
However, in the generic case (for example, when $\|\mathbf C_i\|$ is a random variable with non-zero variance), no faces of the PCH are parallel and no $\mathbf C_i$ belong to the interior of any face of the PCH so the correspondence between vertices of faces of the PCH and species that can coexist we just showed holds.

\section{Analysis of alternative Consumer  Resource Models}
\subsection{Geometric interpretation of consumer resource models with externally-supplied resources}

When considering the externallysupplied resources model, the differential equation for resource dynamics can be written as,
\begin{align}
  \frac{\mathrm dR_\alpha}{\mathrm dt} = R_\alpha\left(\frac{1}{R_\alpha} (\kappa_\alpha - R_\alpha) - \sum_{i=1}^S N_i c_{i\alpha}\right),
\end{align}
for nondepleted resources $\alpha$, allowing for an equation similar to Eq. \ref{Eq:cone-eq}:
\begin{align}
  {\kappa}_\alpha -  R^\star_\alpha
  = 
  \sum_{i,N_i > 0} N_i R^\star_\alpha c_{i\alpha}.
\end{align}
This equation defines the coexistence cones.
The structure of coexisting species, identically to the MCRM, is determined by the sign of $g_i(\mathbf R^\star)$ for each species.
As the form of $g_i$ is the same for the eCRM and MCRM and as $\kappa_\alpha$ being nonnegative implies $R_\alpha$ is nonnegative, the structure of coexistence in the eCRM is identical to that of the MCRM.
That is, coexistence can be enumerated by the faces of the PCH of rescaled consumption vectors, just as in the MCRM.

\begin{figure}[t!]
\includegraphics[width=0.9\linewidth]{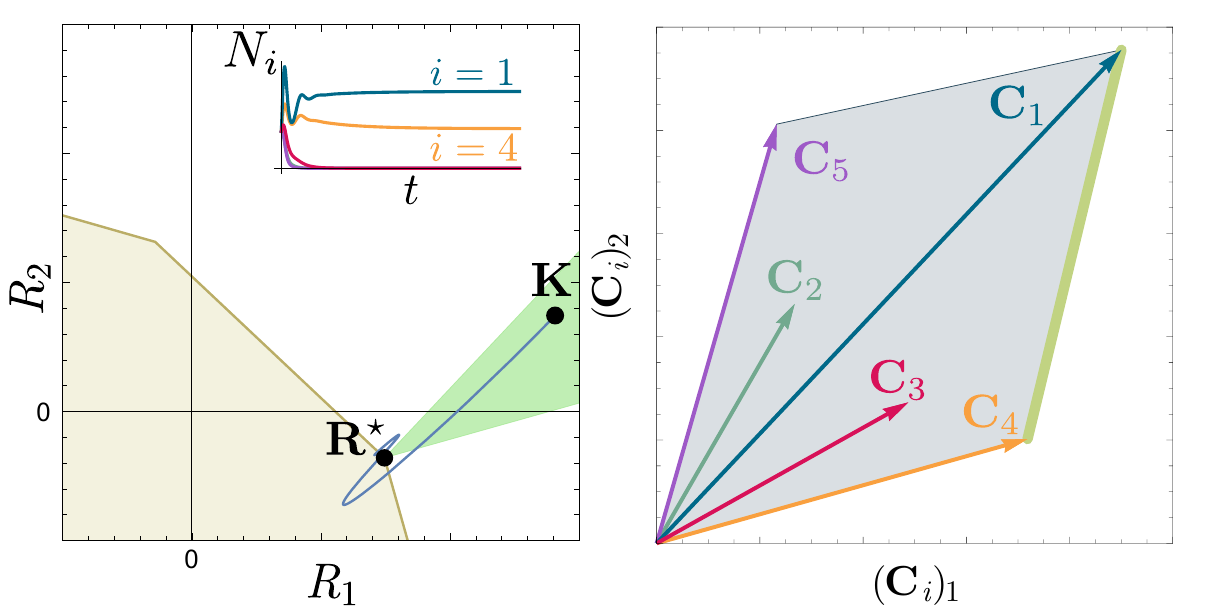}
 \caption{Convex polytope enumerating coexisting species for the Tilman consumer resource model.
  Left: resource-abundance phase space, with $\mathbf K$, $\mathbf R^\star$, and the trajectory $\mathbf R(t)$ plotted as in Fig.~\ref{fig:oldPicture}; here, however, negative values of resource abundances are shown.
  The species population dynamics are additionally inset.
  Right: The relevant convex polytope which is the convex hull of $\{\mathbf C_i\}_{i=1}^S \cup \{\mathbf 0\}$ has faces which enumerate the possible sets of species that can coexist.
  The face corresponding to the coexisting species shown in the left figure is highlighted in green.
  }
  \label{fig:TilmanCH}
\end{figure}

\subsection{Geometric Interpretation of the Tilman consumer resource model\label{appendix:TCRM}}
~

For the TCRM, resource abundances can become nonphysically negative and instead of the subsets of stably coexisting species being enumerated by the faces of positive convex hull, the subsets of stably coexisting species are now enumerated by the faces of the convex hull of $\{\mathbf C_i\}_{i=1}^S \cup \{\mathbf 0\}$, the union of all consumption vectors and the zero vector.
This can be seen by noting that one reason the PCH was introduced is because in the MCRM, resources were required to have nonnegative abundances and could be depleted, but we wanted Eq.~\ref{Eq:cone-eq} to hold in order to make geometric arguments about coexistence cones.
In the TCRM, this equation always holds even if resources are depleted because there is no overall factor of $R_\alpha$ in the TCRM equation for resources [e.g.,~\eqref{Eq:populationTilman}], so it is not necessary to add additional points corresponding to depleted resources when forming the appropriate convex polytope.
The derivation of the properties of this convex polytope is essentially identical to the derivation of the properties of the PCH for the MCRM.
See Fig.~\ref{fig:TilmanCH} for a visualization of the relevant convex polytope and a case where resource abundances become negative.


\bibliography{MCRM-community-geometry_PRE}
\end{document}